# Reducing the metal-graphene contact resistance through laser-induced defects

*Vikas Jangra[1], Satender Kataria[1,2*], Max C. Lemme[1,2*]*

[1]Chair of Electronic Devices, RWTH Aachen University, Otto-Blumenthal-Str. 25, 52074 Aachen, Germany

[2]AMO GmbH, Advanced Microelectronics Center Aachen, Otto-Blumenthal-Str. 25, 52074 Aachen, Germany

**ABSTRACT:**

Graphene has been extensively studied for a variety of electronic and optoelectronic applications. The reported contact resistance between metal and graphene, or rather its specific contact resistance ($R_C$), ranges from a few tens of Ω µm up to a few kΩ µm. Manufacturable solutions for defining ohmic contacts to graphene remain a subject of research. Here, we report a scalable method based on laser irradiation of graphene to reduce the $R_C$ in nickel-contacted devices. A laser with a wavelength of l = 532 nm is used to induce defects at the contact regions, which are monitored *in situ* using micro-Raman spectroscopy. Physical damage is observed using *ex situ* atomic force and scanning electron microscopy. The transfer line method (TLM) is used to extract $R_C$ from back-gated graphene devices with and without laser treatment under ambient and vacuum conditions. A significant reduction in $R_C$ is observed in devices where the contacts are laser irradiated, which scales with the laser power. The lowest $R_C$ of about 250 Ω µm is obtained for the devices irradiated with a laser power of 20 mW, compared to 900 Ω µm for the untreated devices. The reduction is attributed to an increase in defect density, which leads to the formation of crystallite edges and in-plane dangling bonds that enhance the injection of charge carriers from the metal into the graphene. Our work suggests laser irradiation as a scalable technology for $R_C$ reduction in graphene and potentially other two-dimensional materials.

# INTRODUCTION:

Over the past two decades, graphene has been intensively investigated for future applications in electronic and optoelectronic devices, including flexible electronics [1–10]. Metal contacts, like in conventional semiconductor devices, have a profound influence on the performance of graphene-based devices. A low specific contact resistivity ($R_C$) is crucial for any application that requires low power consumption and/or high performance. Graphene has a low density of states near the charge neutrality point (Dirac point), which limits the carrier injection from metals to the graphene, potentially resulting in high $R_C$ [11]. Different strategies have been implemented to minimize $R_C$, which either involve pre-treatments of the metal-graphene (M-G) contact regions to reduce surface contaminations or post-treatments after the devices are fabricated. The pre-treatment category includes methods for cleaning the M-G contact region before metal deposition, such as plasma treatments [12], UV ozone exposure [13], chloroform treatments [14], impingement of $CO_2$ clusters at contact regions to eliminate residues [15], or laser-based cleaning [16]. The post-treatment category includes high temperature - and electrical current annealing techniques to improve interfacial bonding at the M-G contacts. In addition, work function engineering at graphene-metal contacts has been explored with different metals [17] [18], as have double contact geometries to maximize the M-G contact area [19]. Some approaches aimed at reducing $R_C$ by creating edges in the graphene to enable covalent bonds for more efficient carrier injection [20–23]. These strategies were all successful in reducing $R_C$, and it can be derived from the latter that disorder and defects in graphene can favorably influence the chemical and electrical properties of the M-G contact [24–26].

Lasers have long been employed to pattern, deform, and modify materials, including graphene [27–31]. Here, we present a deterministic approach to induce damage in the graphene at the metal contact regions using laser irradiation. We systematically investigate the impact of laser-induced defects on $R_C$ in graphene devices using back-gated transmission line measurements (TLM). A significant, laser-power-dependent reduction in the $R_C$ is noticed within the laser-treated devices.

## 2. EXPERIMENTAL SECTION:

Commercial chemical vapor deposited (CVD) graphene on Copper (Cu) was coated with Polymethyl-methacrylate (PMMA). After removing the Cu in a hydrochloric acid (HCl), hydrogen peroxide ($H_2O_2$), and water solution in a 1:1:10 ratio, the graphene/PMMA stack was transferred onto 2 × 2 cm² silicon chips with a 90 nm thick $SiO_2$ dielectric. The PMMA was removed in acetone [32] (Figure 1a). Graphene channels were patterned using optical lithography and reactive ion etching with oxygen plasma (Figure 1b). The regions where the contact metal was to be deposited were laser irradiated with a wavelength of λ = 532 nm (Figure 1c). We used the laser of our WiTec 300R Raman spectrometer at powers between 13 mW and 20 mW for a duration of 45 sec. A 100x objective was used to focus the laser on the desired locations, resulting in a diffraction-limited spot size of about 300 nm. The laser powers were selected after preliminary tests to produce noticeable damage in graphene, which was tracked by observing irreversible changes in the Raman spectra of graphene (see details in Supporting Information Figure 1). The laser treatment was performed on the graphene channels in a 2 × 3 matrix under each metal contact with a spacing of about 1 µm, controlled by a high-resolution piezo stage for the sample movement. Nickel (Ni) contacts with a thickness of 50 nm were sputter deposited (Figure 1d). An optical image of a TLM structure is shown in Figure 1e. The TLM devices were electrically characterized in a Lakeshore probe station connected to a Keithley 4200 CS parameter analyzer under ambient and vacuum conditions at room temperature.

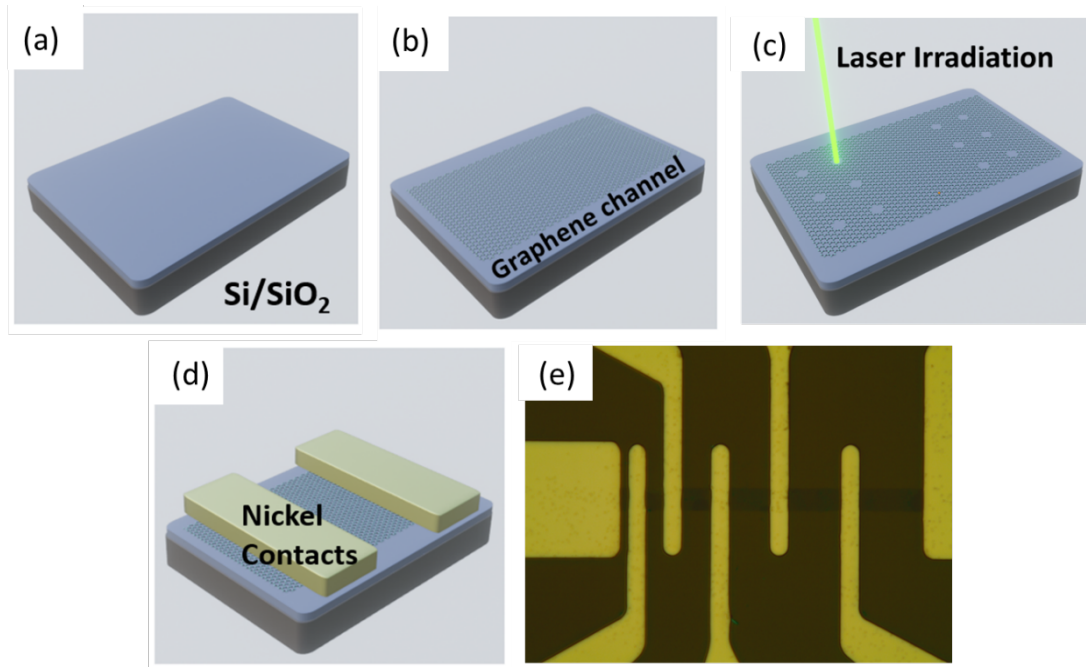

**Figure 1.** Device fabrication: (a)-(d) Schematics showing the process sequence for manufacturing the devices and the laser irradiation of the contact regions in graphene in a matrix of 2 × 3. (e) Optical micrograph of a TLM structure with 3.5 μm channel width with nickel contacts deposited on top of the graphene channel after laser irradiation.

## 3. RESULTS AND DISCUSSIONS:

**3.1. Raman, AFM and SEM characterization.** Raman spectroscopy is a very sensitive technique for identifying and characterizing disorder in graphene and was therefore used to track the effect of the laser irradiation on the graphene [33–36]. Figure 2a shows Raman spectra of graphene acquired just before and after exposure to 15 mW laser power for 45 seconds. This dataset was chosen as a representative example of the power variation to explain the methodology. The spectra were acquired using 2 mW laser power immediately before and after every laser treatment, which was determined to be low enough to avoid permanent damage to the graphene (Supporting Information Figure S1) [37]. All peaks in the spectra were fitted with a Lorentzian function. Before laser treatment, the characteristic G and 2D peaks of crystalline graphene were identified at ~ 1580 cm$^{-1}$ and ~ 2700 cm$^{-1}$, respectively. The G peak corresponds to the $E_{2g}$ optical phonons at the center of the Brillouin zone, while the 2D peak refers to the second-order zone-boundary phonons [33,38]. A very small intrinsic D peak at ~ 1350 cm$^{-1}$ was also observed in graphene before the treatment, which is typical for CVD graphene and is present due to growth or transfer-induced defects [39]. Its intensity is proportional to the amount of disorder in graphene and originates from structural defects due to the second-order Raman scattering process at the boundaries of crystallites [40]. After the laser treatment at 15 mW for 45 seconds, we observed a significant increase in the D peak, indicating the formation of a significant number of dangling bonds and grain boundaries [41]. The intensity ratio of the D peak to the G peak ($I_D/I_G$) is a well-established measure of defect density in graphene and helps to quantify the defect in graphene [35]. Here, the $I_D/I_G$ ratio increases from about 0.028 to 0.5 due to the laser treatment. At the same time, a small shoulder is observed in the G peak, which corresponds to the D' peak at around ~ 1620 cm$^{-1}$. The D' peak appears when significant defects are present in graphene and it originates from an intravalley double resonance process [38,42]. Two other small peaks appear at around ~ 2450 cm$^{-1}$ and ~ 3000 cm$^{-1}$, where the former is a combination of a D phonon and a longitudinal acoustic phonon [43], while the latter corresponds to the combination of D and G bands [35]. We observed similar features in the Raman spectra of graphene in all irradiated regions. Multiple treatments of graphene with 15 mW laser power for 45 sec at different spots show similar $I_D/I_G$ ratios in the range of 0.4 (Figure 2b). Soon after the treatment, high-resolution Raman area mapping was also performed to get an overview of the laser-treated regions. The Raman intensity map of the D peak in Figure 2c shows that the irradiated regions are circular with a diameter of 500 to 600 nm, i.e. larger than the estimated laser spot size of 300 nm. We attribute this to the high energy of about $10^{17}$ W/m² transmitted to the graphene layer, which creates defects that extend beyond the directly irradiated area. This hypothesis is supported by our data, which show larger diameters when we increase the laser power from 13 to 20 mW (Supporting Information Figure S2), as explained in the later section. The line profile of the D-peak intensity across the irradiated regions is extracted from the Raman map in Figure 2c and plotted in Figure 2d. The intensity of the D peak is highest at the center of the laser-treated region and eventually diminishes as we move away from the center toward the untreated graphene (Supporting Information Figure S3).

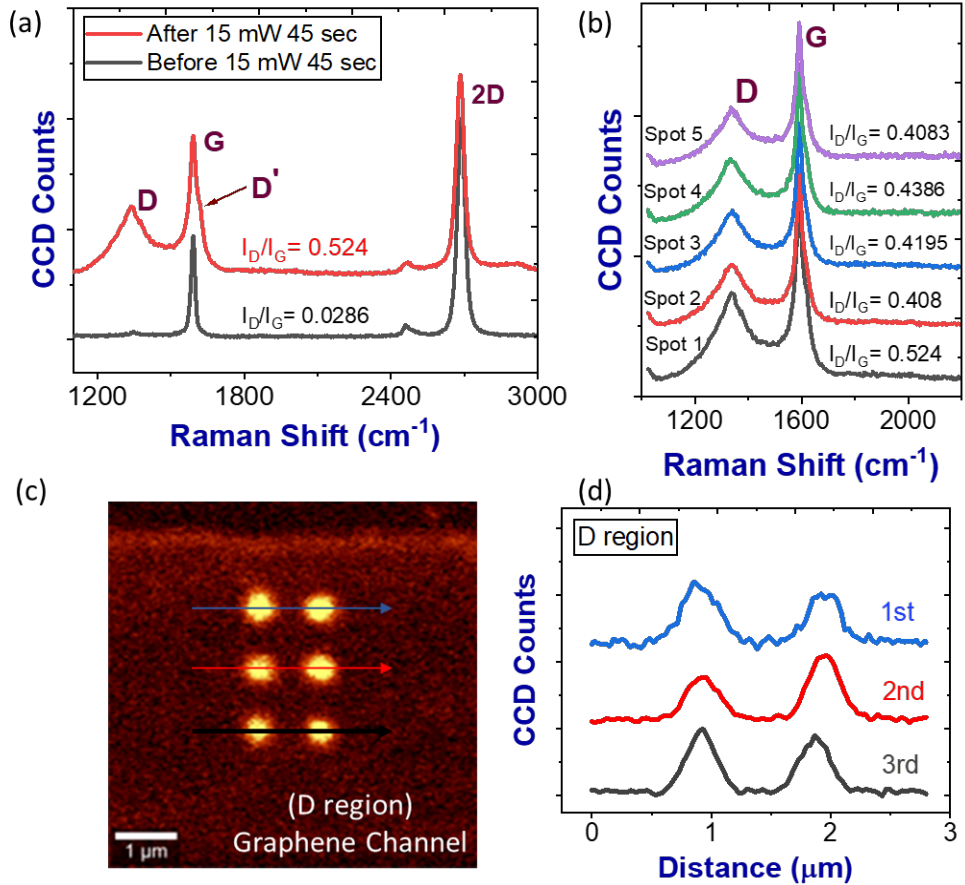

**Figure 2.** Methodology of Raman spectroscopy. (a) Raman spectra of graphene before and after laser irradiation with 15 mW laser power for 45 s. A clear increase in $I_D$ is observed after the treatment, corresponding to an increased defect density in the laser-irradiated region. (b) Raman spectra of five different spots irradiated with identical laser parameters show consistently enhanced D peaks and increased $I_D/I_G$ ratios. (c) Raman map of the D band obtained from laser-irradiated spots within the contact region. The irradiated regions can be clearly identified by their enhanced D-band intensity. (d) Line profiles of the D-band intensities along the three lines indicated in (c).

We further investigated the effect of different laser powers and fabricated another set of devices following the same protocol as described in Figure 1. The contact areas were irradiated with different laser powers of 13, 15, 18, and 20 mW while keeping the exposure time identical at 45 s. Raman measurements were then performed to monitor the laser damage by observing the evolution of the D-band. Typical Raman spectra for each treatment are shown in Figure 3a. The intensity of the D-band indicates an increasing defect generation that scales with the laser power as it is increased from 13 mW to 20 mW, resulting in an enhanced $I_D/I_G$ ratio from ~ 0.29 to ~ 0.70. Raman area maps of the D-bands of the damaged regions illustrate the defect evolution with different laser powers (Supporting Information Figure S2). The lateral sizes of the defect regions in these Raman maps increase from 500 nm to 700 nm with increasing laser power.

Eckman *et al.* observed that regardless of the type of defects generated in graphene, there is always a two-stage evolution of disorder [44]. In stage I, the higher defect density results in more elastic scattering leading to an increase in D-peak intensity, whereas, in stage II, the further increase in defect density leads to an amorphous carbon structure with attenuation of all Raman peaks [44]. Here, all Raman D peaks increase in the laser-treated regions, indicating that the crystalline nature of graphene is still intact (Figure 2). Therefore, the disorder induced by the laser treatment in this work can be classified as stage I. In this stage, the defects induced in graphene are usually quantified by the term "average defect distance ($L_D$)". $L_D$ is > 10 nm at defect concentrations when a laser excitation energy ($E_L$) of 2.33 eV is used for Raman measurements [38]. $L_D$ can be calculated using the following Eq. 1:

$$L_D^2\ (nm^2) = \frac{4.3 \times 10^3}{E_L^4}\left(\frac{I_D}{I_G}\right)^{-1} \quad \text{(Eq 1)}$$

Simultaneously, the defect density ($n_D$) can be calculated as given by Eq. 2 [38].

$$n_D(cm^{-2}) = 7.3 \times 10^9 E_L \left(\frac{I_D}{I_G}\right) \quad \text{(Eq 2)}$$

Figure 3b shows the calculated L$_D$ and n$_D$ values for different laser powers used in the present case, where L$_D$ decreases and n$_D$ increases with increasing laser power.

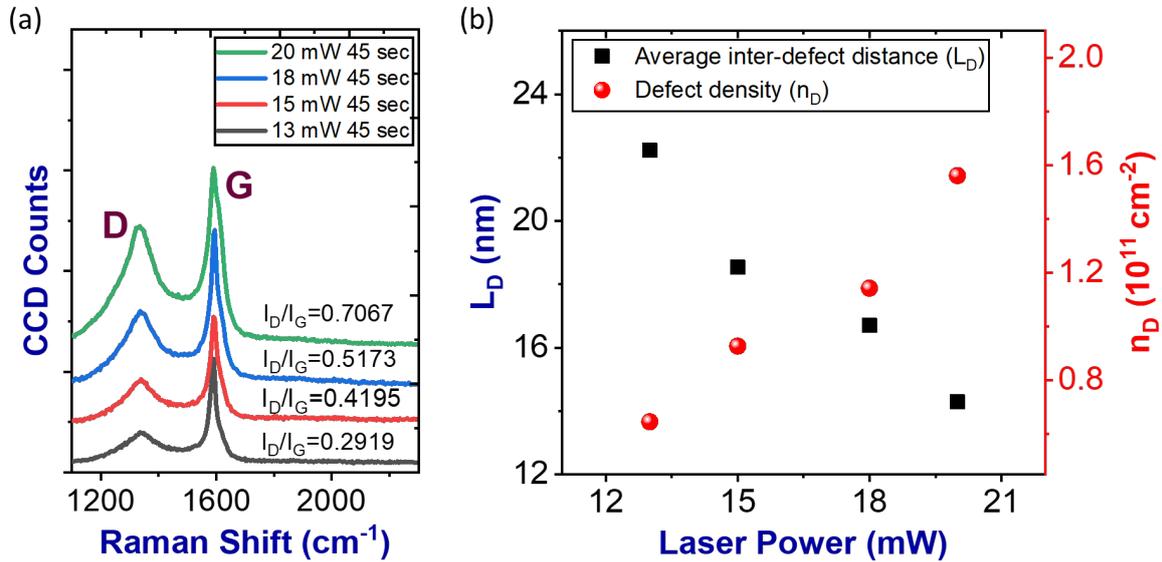

**Figure 3.** Raman analysis of different irradiation powers. (a) Raman spectra of graphene after irradiation with laser powers of 13, 15, 18, and 20 mW for a duration of 45 s. A change in both the D and G peaks is clearly visible, showing that the I$_D$/I$_G$ ratio as a measure of defect density increased drastically (max. 70 %) with increasing laser power, indicating significant defect generation at the contacts. (b) The defect density n$_D$ and the simultaneous representation in terms of average inter-defect distance L$_D$ are shown as a function of increasing laser power. Increased laser power induces more defects in graphene with the maximum n$_D$, and thus the lowest L$_D$, obtained for the 20 mW laser treatment.

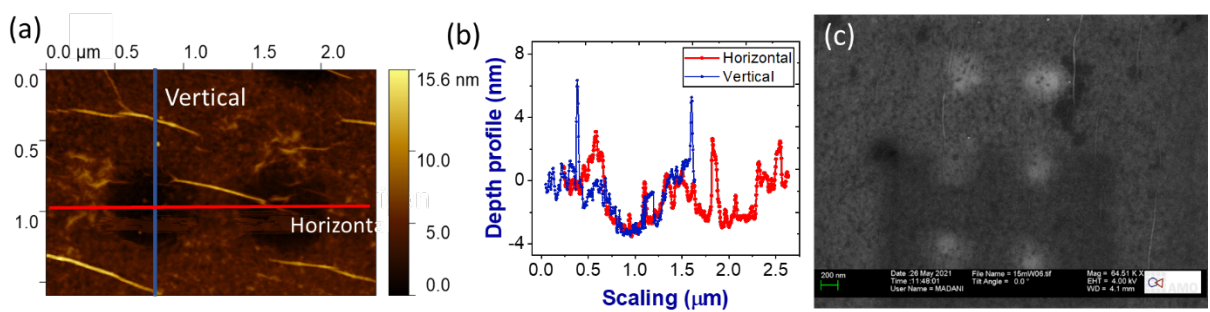

**Figure 4.** Visualization of defects. (a) AFM image of the areas irradiated with 15 mW laser treatment power for 45 seconds. (b) The height profile was extracted from the AFM image in the horizontal and vertical directions along the lines shown in (a). (c) SEM image of the graphene contact area after a 15 mW laser treatment. The irradiated areas appear brighter, indicating lower conductivity due to defects.

Atomic force microscopy (AFM) and SEM analyses were performed to visualize the physical changes caused by the laser irradiation after PMMA removal and graphene channel patterning. The AFM studies reveal visible signs of graphene damage in the laser-irradiated regions (Figure 3a). The regions show the formation of small voids with a depth of 3 – 4 nm, which is attributed to the combined effect of the graphene and resist residue removal. The laser irradiation causes structural changes to the graphene only, while the underlying substrate ($SiO_2$) remains unaffected. This was confirmed by examining the laser-treated spots on graphene and bare $SiO_2$ simultaneously with SEM and AFM (Supporting Information Figure S4). The AFM data confirm the 500 - 600 nm diameter of the affected areas and our interpretation that defects are also generated away from the center of the laser spot (Supporting Information Figure S4). Therefore, we expect a density gradient of graphene edges and dangling bonds from the center of the defect spots to the untreated graphene, consistent with the evolution of the Raman D-peaks in Figure 2c and 2d. SEM images of the 15 mW irradiated samples show bright spots in the graphene at the laser irradiated regions (Figure 4c), corresponding to the enhanced D-peaks in the Raman data. The increased brightness indicates lower conductivity, which can be attributed to more defects in the graphene.

### 3.2. Electrical Measurements.

TLM structures were electrically characterized for samples irradiated with 13,15,18 and 20 mW laser power. Transfer characteristics (drain current $I_d$ versus gate voltage $V_{bg}$) were measured on TLM structures with channel widths of w = 3.5 µm and lengths varying from 3 µm to 11 µm in steps of 2 µm. The drain-source bias voltage was $V_{ds}$ = 100 mV, the back gate voltage was swept from $V_{bg}$ = − 40 V to 60 V, and measurements were performed under both ambient and vacuum conditions (1.3 × 10$^{-4}$ Torr). The measured data for the laser-treated samples is plotted in Figure 5a-d. The data were converted to the total device resistance ($R_T$) for each channel length and plotted against a normalized gate voltage , i.e. $V_{bg}$ - $V_{Dirac}$, to eliminate the effect of residual charge carrier density and to obtain a comparable gate overdrive [45–47] (Figure 5e-h). Transfer curves displaying $I_d$ against $V_{bg}$ and $V_{bg}$ - $V_{Dirac}$ for untreated devices are shown in Supporting Information Figure S5. The contact resistance ($r_C$) was extracted by extrapolating $R_T$ with the channel length (L) at different induced carrier densities ($n_i$), according to the TLM method [45] using Eq. 3, where $R_{sh}$ denotes the sheet resistance of the channel (see Supporting Information Figure S6 for the linear fitting of $R_T$ and L).

$$R_T = \frac{R_{sh}}{W}L + 2r_C \qquad (Eq\ 3)$$

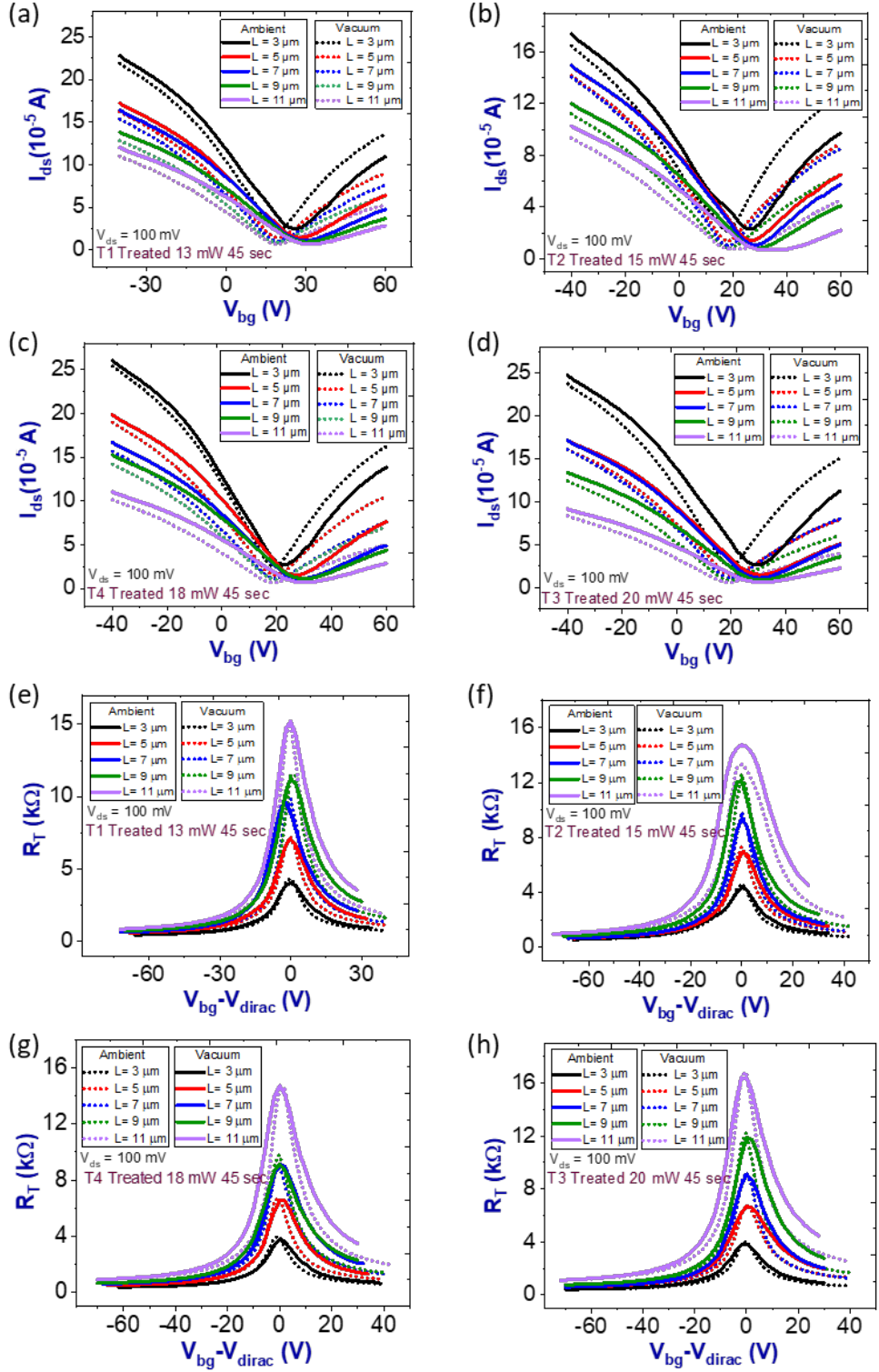

Figure 5. (a)-(d) Back-gated TLM structures were measured to obtain the transfer characteristics of laser-treated devices (13,15,18 and 20 mW - 45 s) under ambient and vacuum conditions. (e)-(h) The total resistance ($R_T$), plotted as a function of the normalized back-gate potential ($V_{bg} - V_{Dirac}$) at a drain-source bias of $V_{ds}$ = 100 mV for different channel

lengths (L). The maximum $R_T$ is obtained for the largest L = 11 μm. These characteristics provide the contact resistance ($r_C$) as a function of the induced carrier density from a single transfer curve.

The extracted $r_C$ values were then converted to the specific contact resistivity $R_C$ (Ω μm), which takes into account the width of the contact. This is important when studying 2D materials. Therefore, the $R_C$ values were obtained by multiplying $r_C$ by the channel width w = 3.5 μm, which is consistent for all devices. They are compared for the differently treated TLM devices in Figure 6 for different gate overdrives $V_{bg}$ - $V_{Dirac}$. However, we omit the values obtained for $V_{bg}$-$V_{Dirac}$ > -20 V, because the $R_C$ values were fluctuating and sometimes negative, a known unphysical result [47]. We also evaluated $R_C$ as a function of the induced charged carrier density ($n_i$), which is plotted on the second x-axis of the data in Figure 6, based on Eq. 4 [45]. Here, $C_g$ denotes the oxide capacitance calculated using Eq. 5 where ε and $ε_o$ are the relative and absolute dielectric constants and an oxide thickness of $t_{ox}$ = 90 nm $SiO_2$.

$$n_i = C_g |V_{bg} - V_{Dirac}| \qquad (Eq\ 4)$$

$$C_g = εε_o/t_{ox} \qquad (Eq\ 5)$$

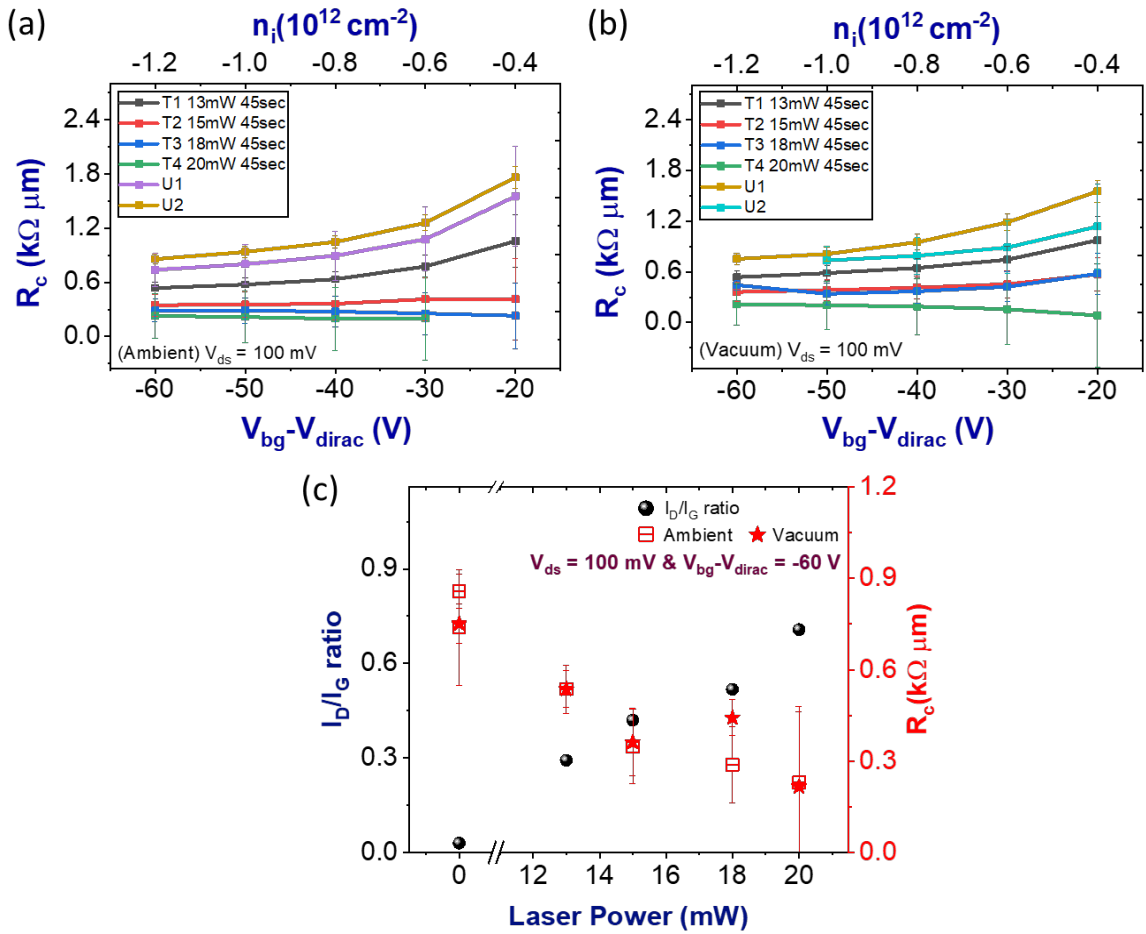

**Figure 6.** (a),(b) Specific contact resistivity ($R_C$) extracted for 13, 15, 18 and 20 mW laser-treated and untreated devices represented as a function of $V_{bg} - V_{Dirac}$ or induced charge carrier density ($n_i$) in ambient and vacuum at $V_{ds}$ = 100 mV. Reduced $R_C$ for the laser-treated devices is clearly witnessed. The legends represent the device nomenclature used while fabricating the device (T = treated with laser and U = untreated). The vertical error bars indicate the standard deviation during $R_C$ extraction. Simultaneously, an overall effect of different laser powers on defect generation is shown in Figure 6c in terms of the $I_D/I_G$ ratio obtained from Raman measurements at the metal-graphene contact regions and its effects on $R_C$.

Figure 6a and 6b show the extracted $R_C$ values from different devices under ambient and vacuum conditions, respectively. $R_C$ was significantly lower in laser-treated devices compared to untreated devices in both ambient (Figure 6a) and vacuum conditions (Figure 6b). $R_C$ was consistent in the range of 700 Ω μm – 900 Ω μm for the untreated devices but showed a lower value of about 500 Ω μm for the laser-treated device (13 mW) at a carrier density of - $1.2 \times 10^{12}$/cm$^2$ (both in ambient and vacuum). The $R_C$ value decreased further for the devices treated with higher laser powers, and the minimum $R_C$ was decreased to about 250 Ω μm for the 20 mW treated device. The relative change in $R_C$ is more pronounced at lower $n_i$, reaching a decrease of up to 60 % at a carrier density of - $0.4 \times 10^{12}$ in treated devices. Similar observations were made for a reduction in $R_C$ of up to 50 – 60 % at the highest induced carrier densities measured under vacuum conditions, as shown in Fig. 6b. This reduction can be attributed to the fact that higher laser power induces higher defect density in the exposed region, as indicated by an increase in the $I_D/I_G$ ratio (Figure 3a). These defects provide a higher probability for in-plane covalent bonding between the contact metal and graphene. Removing polymer residues has been suggested as an alternative reason for the $R_C$ decrease, for example using a laser power of 10 mW for 300 s [16]. However, we included another lithography step after the laser treatment to deposit metal contacts. Thus, another underlying layer of polymer residues was inadvertently present in the patterned areas, producing the same effect as we observe in untreated devices. This signifies that the $R_C$ reduction is dominantly caused by the defects induced in the graphene at the metal-graphene interface. Figure 6c summarizes the correlation between laser power and defect density and its effect on the $R_C$. The minimum $R_C$ of 250 Ω μm was obtained for the 20 mW laser-treated devices at the highest charge carrier density of - 1.2 x 10$^{12}$ cm$^{-2}$, corresponding to a decrease of about 70 – 80 % compared to the untreated devices. Table 1 compares the reported $R_C$ values obtained with different methods using engineering and patterning of the contact regions. The comparison is not straightforward because the studies used different metals and extracted the data at different charge carrier densities. Among these methods, laser irradiation has the potential to significantly reduce the $R_C$ of graphene devices with a high degree of simplicity and scalability, making it highly applicable.

**Table 1.** Comparison of different methodologies and the lowest values of $R_C$ achieved therein.

| Reference | $V_{bg}$-$V_{Dirac}$ (V) | Lowest $R_C$ (Ω µm) | Contact Metal | Methodology /Complexity | % decrease w.r.t untreated devices |
|---|---|---|---|---|---|
| [21] | Dirac point | 23 | Au | E-beam holey contacts/High | 88 |
| [19] | -20 | 200 | Ti/Pd – Ti/Pd/Ni | Double contact geometry/Medium | 62 |
| [22] | Dirac point | 45 | Au | E-beam holey contacts/High | 91 |
| [20] | - | 125 | Cu | Patterned contacts/Annealing/Medium | 32 |
| [48] | -40 | 207 | Pd | Different contact metal/Low | - |
| This work | - 60 | 250 | Ni | Laser irradiation/Low | 70 |

## 4. CONCLUSIONS

We have used a visible green laser (532 nm) with different powers from 13 mW to 20 mW to induce defects in graphene at metal-graphene contact regions and studied the effect of the laser irradiation on the specific contact resistivity $R_C$. The defects were measured with Raman spectroscopy by tracking the D-band intensity and the $I_D/I_G$ ratio. The defect density increases systematically with the laser power, reaching about $1.6 \times 10^{11}$ cm$^{-2}$ at 20 mW. SEM and AFM studies revealed physically damaged regions suggesting the presence of dangling bonds and crystallite edges. Back-gated TLM structures with Ni contacts were used to extract $R_C$ with and without laser irradiation. The laser-treated devices exhibit a lower $R_C$ compared to the untreated ones, which scales with laser power. The lowest $R_C$ of 250 Ω μm was obtained at the highest laser power of 20 mW, a 70 % reduction compared to untreated devices. This is attributed to the enhanced in-plane charge carrier injection from metal to graphene through covalent metal-carbon chemical bonds. The proposed method can be easily scaled, implemented, and automated to engineer the $R_C$ in graphene and potentially other 2D material-based devices.

## ASSOCIATED CONTENT

### Supporting Information

Preliminary tests for optimizing laser powers and time, Raman and SEM mapping of defect induced regions and simultaneous effect on $SiO_2$ substrate, Raman scan of defects spread within the graphene channel, transfer curves for untreated devices in ambient and vacuum and linear fitting of total resistance and channel length to extract specific contact resistivity.

## AUTHOR INFORMATION

### Corresponding Authors

**S. Kataria** – Chair of Electronic Devices, RWTH Aachen University, Otto-Blumenthal-Str. 25, 52074 Aachen, Germany and AMO GmbH, Advanced Microelectronics Center Aachen, Otto-Blumenthal-Str. 25, 52074 Aachen, Germany; https://orcid.org/0000-0003-2573-250X; Email: kataria@amo.de

**Max C. Lemme** – Chair of Electronic Devices, RWTH Aachen University, Otto-Blumenthal-Str. 25, 52074 Aachen, Germany and AMO GmbH, Advanced Microelectronics Center Aachen, Otto-Blumenthal-Str. 25, 52074 Aachen, Germany; https://orcid.org/0000-0003-4552-2411; Email: lemme@amo.de

### Author
**V. Jangra** - Chair of Electronic Devices, RWTH Aachen University, Otto-Blumenthal-Str. 25, 52074 Aachen, Germany; https://orcid.org/0000-0002-1339-7137

### Author Contributions

S.K. and M.C.L. conceived the experiments. V.J. prepared the samples and performed the experiments. V.J. and S.K. analyzed the data. The manuscript was written by V.J. and revised by S.K. and M.C.L. The final version of the manuscript was edited and approved by all authors.

## ACKNOWLEDGEMENTS:

We acknowledge funding by the German Ministry of Education and Research (BMBF) through the project GIMMIK (03XP0210) and the European Union's Horizon 2020 research and innovation programme through the project Graphene Flagship Core 3 (881603). Vikas Jangra acknowledges the DAAD KOSPIE fellowship for pursuing the research work.

Supporting Information

# Reducing the metal-graphene contact resistance through laser-induced defects

*V. Jangra[1], S. Kataria[1,2,*], M. C. Lemme[1,2,*]*

[1]Chair of Electronic Devices, RWTH Aachen University, Otto-Blumenthal-Str. 2, 52074 Aachen, Germany

[2]AMO GmbH, Advanced Microelectronics Center Aachen, Otto-Blumenthal-Str. 25, 52074 Aachen, Germany

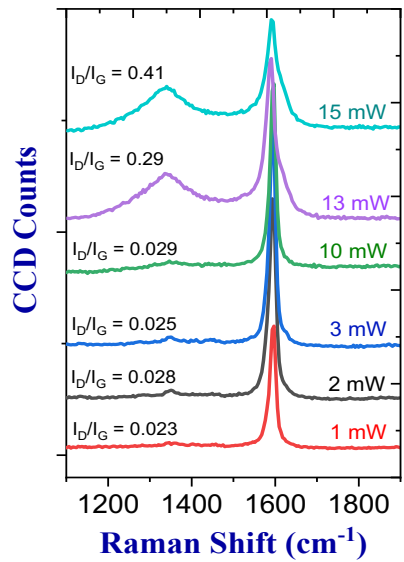

**Figure S1:** Shown above are the Raman measurements of preliminary tests made to induce defects in graphene. All the measurements were carried out using 2mW laser power to efficiently extract the characteristic peaks of graphene to make sure no damage was induced during characterization. We obtained sufficient defects induced at around 13 mW and thereafter 15, 18, and 20 mW laser powers were also chosen in this work to make a comparison.

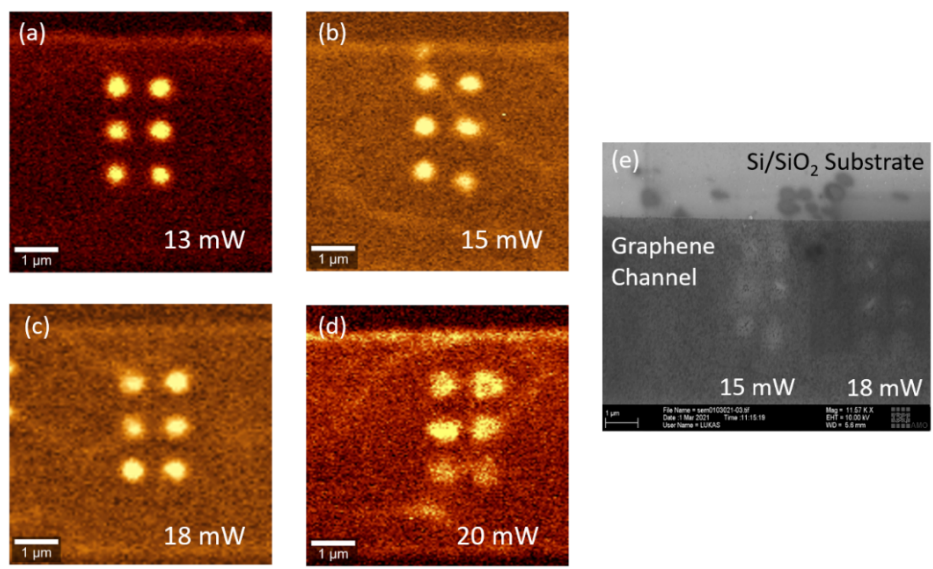

**Figure S2:** Raman mapping of D region of laser-treated graphene channel with different laser powers (13,15,18 and 20 mW) indicating generation of defects of around 500-700 nm in diameter (a)-(d). Simultaneously shown is the SEM (4 keV) image of 15 and 18 mW laser treatment on graphene in a 2x3 matrix displayed with the bright parts indicating induced physical changes (e).

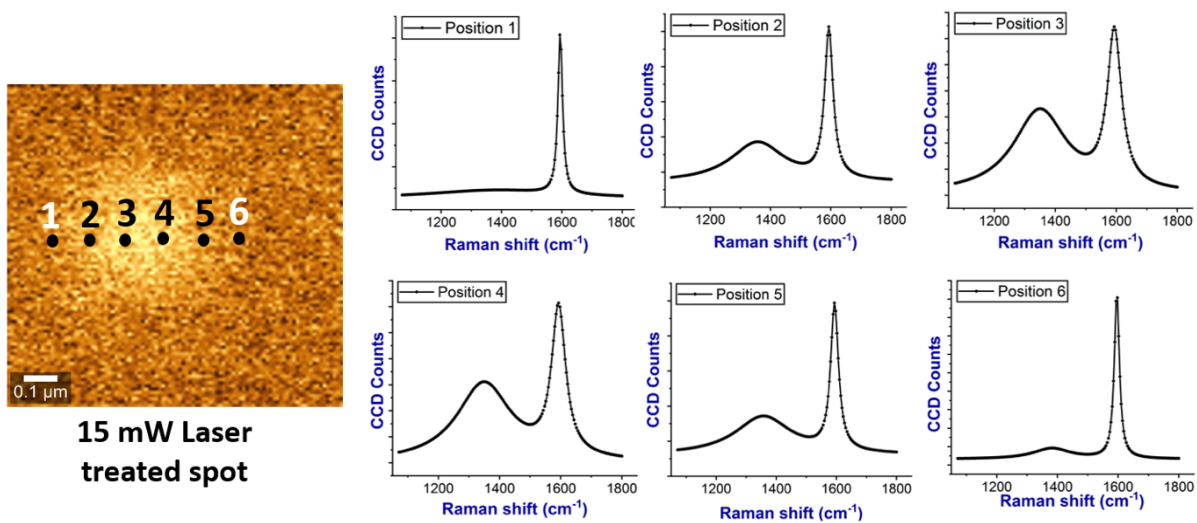

**Figure S3:** High-resolution Raman scan of 15 mW laser treatment on graphene indicating a defective region of higher brightness. The Raman spectra with D and G peak intensities at each numbered location show higher D peaks at the laser-treated center and decreasing D peaks as the location moves towards untreated graphene.

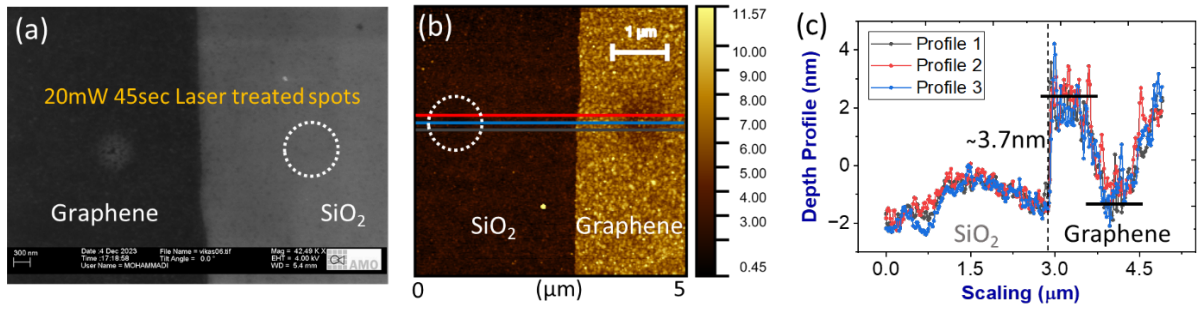

**Figure S4:** (a) SEM image of 20 mW and 45 sec irradiated spots on graphene and SiO$_2$ substrate (highlighted with circle). (b) AFM image of the corresponding spots shows graphene removal, but no effect is visible on the SiO$_2$ substrate. (c) The line profile indicates resist residue removal from the graphene of about ~ 3.7 μm. The data show that the substrate was unaffected by the laser irradiation, as there were no discernible structural changes in the SEM or AFM images.

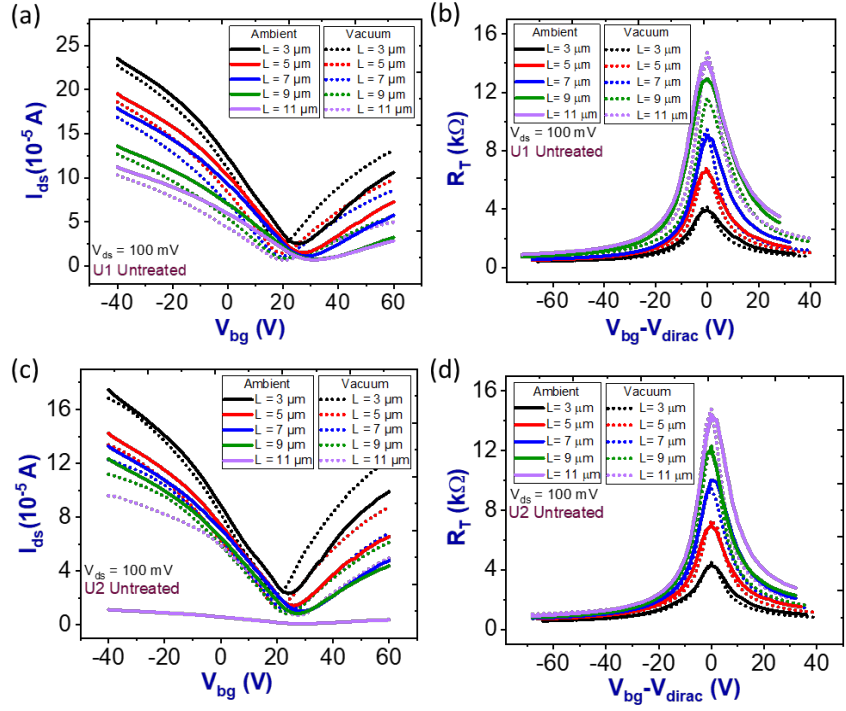

**Figure S5:** Transfer curves of untreated device and laser treated (13, 18, and 20 mW) in ambient. The obtained curves at different $V_{bg}$ display shifts in the Dirac point, indicating an n-doped graphene channel. The shift is noticed less in the case of vacuum measurements for the same devices due to the decreased environmental adsorbates on the graphene channel. The $I_{ds}$ – $V_{bg}$ curves were converted to $R_T$ – $V_{bg}$ - $V_{Dirac}$ curves to calculate specific contact resistivity ($R_C$) using Eq. 3.

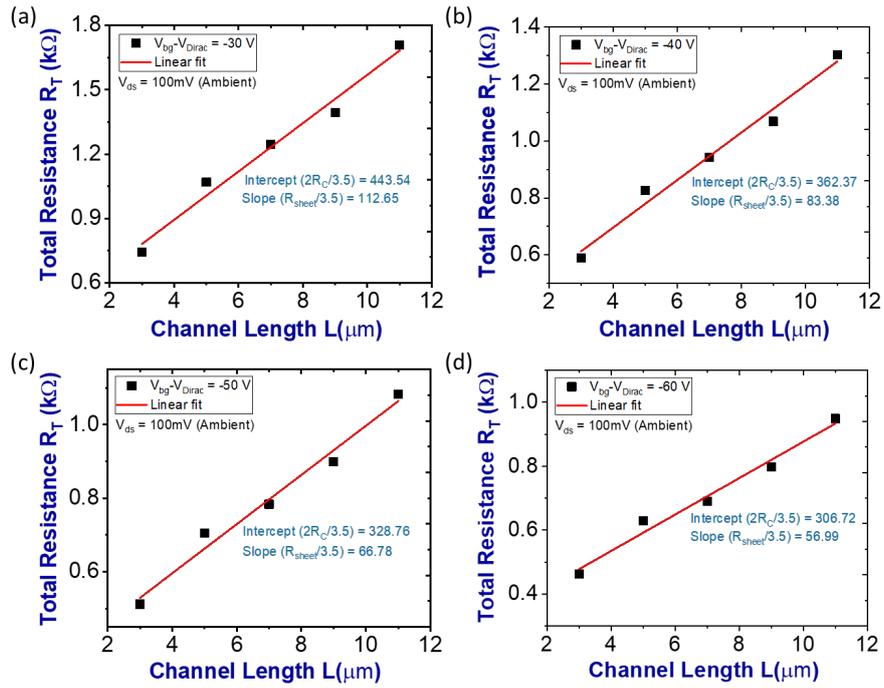

**Figure S6:** Examples of linear fitting of total resistance ($R_T$) vs. channel length (L) for a 13 mW and 45-sec laser-treated device at different $V_{bg}$ - $V_{Dirac}$ for a fixed $V_{ds}$ = 100 mV. 2x$R_C$ can be extracted from the intercept of the fit with the y-axis. Similar plots were made to extract $R_C$ for all laser-treated and untreated devices.